\documentclass[a4paper,fleqn]{cas-sc}

\usepackage{multirow}
\usepackage{diagbox}
\usepackage{tabularx}
\usepackage[square,numbers]{natbib}
\def\tsc#1{\csdef{#1}{\textsc{\lowercase{#1}}\xspace}}
\tsc{WGM}
\tsc{QE}
\tsc{EP}
\tsc{PMS}
\tsc{BEC}
\tsc{DE}

\newcommand{\aadl}{\textsf{\em AADL Paper} }

\begin{document}
\let\WriteBookmarks\relax
\def\floatpagepagefraction{1}
\def\textpagefraction{.001}

\shorttitle{}

\shortauthors{Minghui Sun et~al.}

\title [mode = title]{Characterizing the Identity of Model-based Safety Assessment: A
Systematic Analysis}

%

\author[1]{Minghui Sun}[type=,
                        auid=000,bioid=1,
                        prefix=,
                        role=,
                        orcid=0000-0002-9220-1576]

\cormark[1]

\ead{minghuis@iastate.edu}

\ead[url]{}

\credit{}


\address[1]{Iowa State University, Ames, 50011, Iowa, USA}


\author[2]{Smitha Gautham}[style=]
\ead{gauthamsm@vcu.edu}

\author[2]{Carl Elks}[style=]
\ead{crelks@vcu.edu}

    
\address[2]{Virginia Commonwealth University, Richmond, 23284, Virginia, USA}

\author[1]{Cody Fleming}[style=]
\ead{flemingc@iastate.edu}

\cortext[cor1]{Corresponding author}


\begin{abstract}
Model-based safety assessment has been one of the leading research thrusts of the System Safety Engineering community for over two decades. However, there is still a lack of consensus on what MBSA is. The ambiguity in the identity of MBSA impedes the advancement of MBSA as an active research area. For this reason, this paper aims to investigate the identity of MBSA to help achieve a consensus across the community. Towards this end, we first reason about the core activities that an MBSA approach must conduct. Second, we characterize the core patterns in which the core activities must be conducted for an approach to be considered MBSA. Finally, a recently published MBSA paper is reviewed to test the effectiveness of our characterization of MBSA. 
\end{abstract}


\maketitle

\section{Introduction}\label{sec:intro}
Model-based safety assessment\footnote{Sometimes also known as model-based safety analysis.} (MBSA) has been around for over two decades. The benefits of MBSA have been well-documented in the literature, such as tackling complexity, introducing formal methods to eliminate the ambiguity in the traditional safety analysis, using automation to replace the error-prone manual safety modeling process and ensuring the consistency between the design model and the safety model \cite{prosvirnova2014altarica}.  We have seen a flourish of research development over the years.
Prominent modeling languages from the academic realm, such as AADL-EMV2 \cite{larson2013illustrating,procter2020aadl,feiler2017automated}, AltaRica \cite{machin2019modeling, bieber2004safety, mortada2014safety, tlig2018autonomous}, and HipHops \cite{kabir2019conceptual, chen2013systems} are generally considered MBSA \cite{bozzano2015safety}. In addition, there are a growing number of commercial tools aimed at applying general model-based approaches to safety analysis, such as SCADE \cite{scade}, CAMET \cite{camet}, MADe \cite{made}, and Medini \cite{medini}. 
However, MBSA approaches are significantly dissimilar in the literature \cite{stewart2021aadl}. A significant question remains: what exactly is MBSA?

There is a lack of consensus on the identity of MBSA. For example, is Formal Methods applying to safety analysis MBSA (e.g., \cite{desgeorges2021formalism})? Are MBSA approaches different from the general model-driven approaches applied to safety-critical systems (e.g., \cite{pajic2012model-driven, althoff2010reachability}), and if so, how? It is usually claimed that MBSA contributes to the current literature by, for example, handling the increasing complexity of safety-critical system, integrating different views, finding design flaws early, reusing previously developed artifacts, introducing automation to reduce time and cost, and structuring unstructured information.
However, these contributions can also be found in the general Model-based Design community \cite{scippacercola2016,akdur2018survey,liebel2014assessing,panesar-walawege2013supporting,broy2012what,barbieri2014model-based,jayakumar2020systematic}. Obviously, MBSA is closely related to all the communities above, but a question such as what makes an approach MBSA is still left unanswered. 

This ambiguity has significant implications. First, it jeopardizes the identity of MBSA as a main research thrust of the system safety engineering community. Without a clear definition, MBSA can quickly become a buzzword that any discipline can claim as long as the work uses computer models and has safety implications. 
Such an identity crisis impedes MBSA researchers communicating to a broader audience (even to some safety practitioners), promoting the significance of MBSA, and forging new breakthroughs and synergies with the rest of the scientific communities.

Furthermore, because of the lack of consensus on the identity of MBSA, some people equate MBSA with those high-profile modeling languages such as AADL-EMV2 and AltaRica, or the development of new modeling languages and tools similar to them. It is true that those modeling languages are crucial for the success of MBSA, but one should not forget MBSA ultimately serves the overall goal of safety assurance. Obviously, MBSA alone is insufficient to achieve this goal. Narrowly focusing on developing new modeling languages or applying the current modeling languages to new domains may miss other equally significant issues and research opportunities to push MBSA forward. For MBSA to be used effectively for safety assurance, it is imperative to address key gaps and challenges that the safety community must address. However, before identifying these challenges, we must first define MBSA.

Therefore, to help the safety community to eventually reach a consensus on the identity of MBSA, this paper reasons and characterizes MBSA by answering the following two questions:
\begin{itemize}
    \item Question 1: What is the minimal set of activities that must be conducted for an approach to be considered MBSA? We call them the ``\textbf{core activities}'' of MBSA.
    \item Question 2: What is the minimal set of patterns in which the core activities must be conducted for an approach to be considered MBSA? We call them the ``\textbf{core patterns}'' of MBSA.  
\end{itemize}

\paragraph{Research method.}
A common approach to investigating the identity of a research subject is literature review. 
In a conventional review, papers are systematically searched and then screened based on a set of pre-defined criteria with respect to the scope of the subject matter. However, we do not have the privilege, because MBSA is ill-defined. The results of such a review will be inevitably biased towards the criteria selected. Then, to achieve a consensus, one will have to compare one set of criteria against another, which will eventually require an understanding of the reasoning behind the different sets of criteria. Compared with the specific criteria selected to characterize MBSA, a more fundamental question is how to reason about MBSA.

Therefore, our approach provides a way to reason about MBSA. We start from a broader and better-established scope than MBSA to reason about the ``general activities'' that may be conducted by an MBSA approach. After that, we refine this scope into a minimal set of well-understood activities that an MBSA approach must absolutely conduct (called the ``core activities''). Then, we characterize the ``general patterns'' in which the core activities are conducted in the literature. Finally, we summarize a minimal set of common characteristics of the general patterns (called the ``core patterns'') that an approach must exhibit to be considered MBSA.

Note that we do not have a ``Literature Review'' section in this paper because this entire paper is presented based on reviewing and analyzing the MBSA literature. Furthermore, although almost all MBSA literature mention their interpretations of MBSA, these interpretations usually take an implicit stance that characterizes MBSA only to emphasize the technical features of the particular paper. 
To the best of our knowledge, there is no comprehensive study that is specifically dedicated to the investigation of the identity of MBSA.

\section{The activities of MBSA}\label{sec:background}
In this section, we first identify a set of general activities of MBSA. Then, we reason about a minimal set of core activities that an MBSA approach must conduct. 
\subsection{Model-based Design (MBD)}
We start by putting MBSA in a broad scope:
\begin{center}
\emph{Assertion 1: MBSA is Model-based Design applied to system safety assessment.}
\end{center}
\noindent Although it is difficult to find direct evidence to support this assertion because there is no consensus on an exact definition of MBSA, this assertion is consistent with \cite{joshi2006model-based,abdellatif2020model} that claim MBSA is an extension of safety analysis to Model Based Design. 
More importantly, our confidence in starting from this assertion is in its broadness, meaning we will not miss important characteristics of MBSA by starting off with a too narrow view.

\begin{figure}[ht!]
\centering
\includegraphics[width=0.6\linewidth]{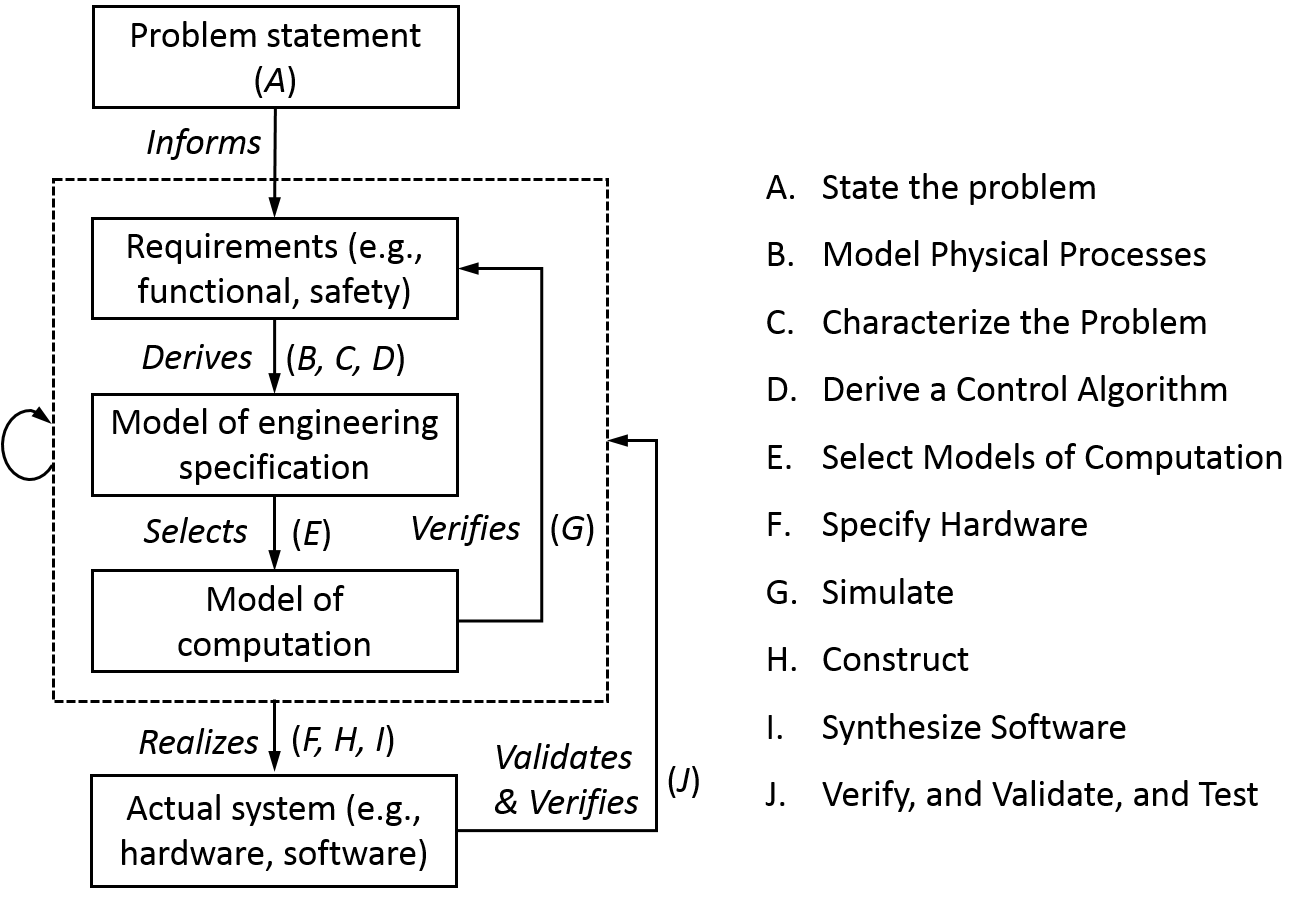}
    \caption{The MBD process. The arrows of the left can be mapped to the steps to the right (adopted from \cite{jensen2011model-based}).}
    \label{fig:mbd_process}
\end{figure}

We are aware that MBD is an overloaded term. Numerous papers tried to differentiate it from ``model-driven engineering'', ``model-driven design,'' and ``model-driven architecture'' \cite{whittle2013state,scippacercola2016}. It is not our intention to define these terms. However, we need a clear understanding of MBD to start the reasoning of MBSA. Hence, we adopt the process proposed by \cite{jensen2011model-based} as the ``ground truth'' of our discussion. The left of the Fig.\ref{fig:mbd_process} is the proposed MBD process which can be mapped to the detailed 10 steps of \cite{jensen2011model-based} at the right. We explain all the involved steps in this section. 

\paragraph{The inner loop (Step A-G).} An MBD process starts with a problem statement (Step A) which details the goal of the design activities. The problem statement then informs the definition of requirements such as functional requirements and safety requirements. 
Second, models are derived to solve the stated problem, which corresponds to Step B, C and D in \cite{jensen2011model-based}. 
The resulting models are a representation of the engineering specifications that will be used to realize the actual system, hence are called the ``model of engineering specification''.  
Third, a modeling language is selected to further represent the aspects under study of the model of engineering specification such as functional, physical, safety and environment \cite{gautham2022model}, leading to an executable model (called ``model of computation'', Step E) that can be automatically computed and verified against the requirements. 
Fourth, Step G closes the inner loop by automatically verifying the model of computation against the requirements. If the verification is passed, the MBD process proceeds to the outer loop; otherwise the model of engineering specification must be updated until the model of computation satisfies the requirements.

\paragraph{The outer loop (Step F-J).} Given the model of computation, specifications are derived (Step F); hardware is constructed in accordance with the specification (Step H); software is developed or automatically generated from the model and synthesized with the hardware (Step I).
Finally, Step J closes the outer loop. The prototype or the actual system is built to verify against the requirements, and eventually the resulting system is validated against the problem statement made at the beginning of the MBD process. 

\vspace{2mm}
As stated in Assertion 1, MBSA is MBD applied to safety assessment. Safety assessment can happen during both the inner loop and the outer loop of MBD. However, safety assessment often emphasizes capturing design problems early before the software and the hardware are built, at which time correcting a design error can be costly or even impossible \cite{fleming2015safety}. 
We acknowledge that model-based approach can still be useful at a later stage (e.g.,\cite{henderson2016model}), but MBSA is nevertheless still most effective at the early stages of design. 
Therefore, we refine the scope of MBSA from Assertion 1 to Assertion 2. 
\begin{center}
\emph{Assertion 2: MBSA is the inner loop of MBD applied to safety assessment.}
\end{center}

\subsection{The general activities of MBSA}\label{sec:activity}
\begin{figure}[ht!]
    \centering   \includegraphics[width=0.55\linewidth]{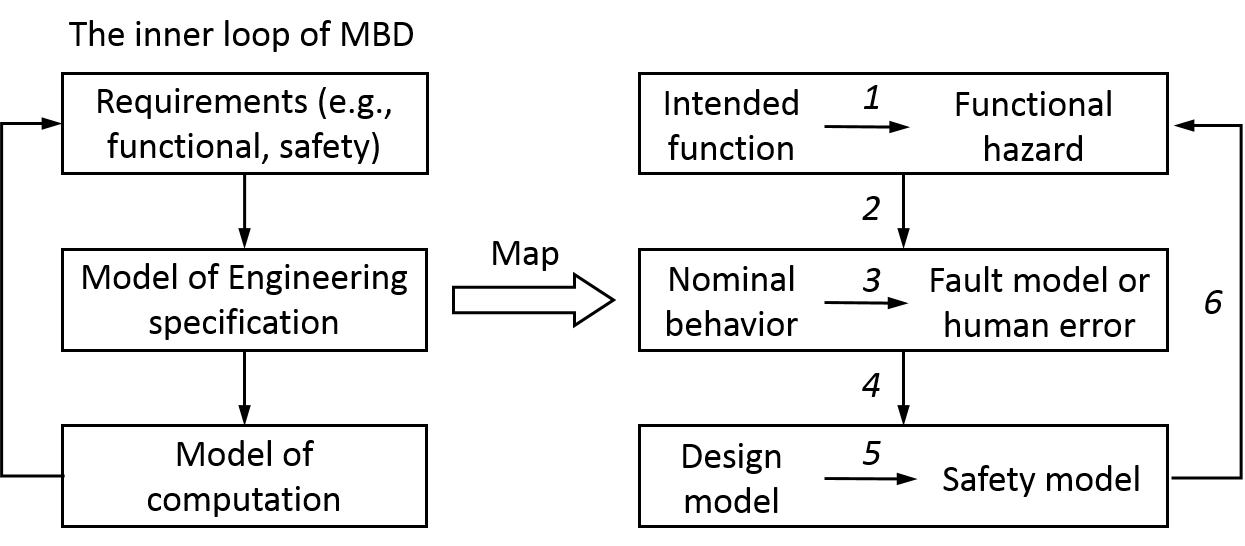}
    \caption{A holistic view of safety assessment in the context of the inner loop of MBD. Note that the generation of the intended function, the nominal behavior, and the design model are not the tasks of the safety assessment process, but they support the safety assessment process.}
    \label{fig:mbdsafety}
\end{figure}

Next, we take a holistic view of safety assessment by applying the inner loop of MBD to the safety assessment process (Fig.\ref{fig:mbdsafety}). First, the ``requirements'' are the intended function to achieve and the associated functional hazard to avoid. Second, the ``model of engineering specification'' is the nominal behavior to achieve the intended function, and the component fault or human error that may lead to the functional hazard. 
Third, the ``model of computation'' is the design model and the safety model to be analyzed to ensure the safety risk associated with the functional hazard is acceptable.   

We explain in detail each activity of the safety assessment process (i.e., the six arrows in Fig.\ref{fig:mbdsafety}).
\paragraph{Functional hazard identification (Arrow 1).}
Given the intended functions, Arrow 1 identifies the functional hazards. Traditional approaches for hazard identification include, for example, Hazard and Operability Analysis \cite{crawley2015hazop} and Functional Hazard Assessment \cite{wilkinson1998functional}.
More recently, the model-based approaches are also applied to (partially) automate the hazard identification process \cite{habli2009functional, beckers2013structured,cancila2009sophia,guiochet2016hazard,johannessen2001hazard,kaleeswaran2019domain,maitrehenry2011toward,jiang2020mbse,haider2020mbsa,savelev2021fmea}.

\paragraph{Causal factors identification (Arrow 2).}
After the nominal behavior is defined by the design team, Arrow 2 identifies the casual factors that may lead to the hazard. There are two types of casual factors: design errors of the nominal behaviors, and the component fault and the human error that may lead to the off-nominal behaviors. Such an activity is usually called deductive safety analysis \cite{vilela2017integration},  sometimes also known as top-down \cite{grigoleit2016qsafe,chaari2016transformation} or effect-to-cause \cite{fenelon1994towards} safety analysis.  
For example, two well-known deductive safety techniques are Fault Tree Analysis \cite{ruijters2015fault} and Systems Theoretic Process Analysis \cite{leveson2016engineering}. Note that the identified design errors have to be corrected before the design can proceed to the next step. Therefore, only the component fault and the human errors flow to the next step.

\paragraph{Component fault and human error characterization (Arrow 3).}
Arrow 3 characterizes the component faults and human errors identified by Arrow 2. This activity requires scientific understanding about how a component fails and/or how a person makes mistakes. Studies are available in the system safety literature on how components fail such as \cite{joshi2007behavioral,wille2019contributions,wolforth2010capture, guangyan2010system,o2014early,procter2020aadl}.
Similar efforts can also be found in the human factors literature on human reliability analysis such as \cite{mosleh2004model,french2011human}.  

\paragraph{Modeling language selection (Arrow 4).}
Arrow 4 selects the modeling language for the safety model. We are aware that there is a tight coupling between the modeling languages for the design model and the safety model. But we mainly focus on the modeling language for the safety model in this paper.

\paragraph{Safety model construction (Arrow 5).}
Arrow 5 builds the safety model based on the design model. Although the individual component faults are already characterized by Arrow 3, they still need to be weaved together so that the resulting safety model can describe the off-nominal behavior of the system precisely.

\paragraph{Safety analysis (Arrow 6).}
Arrow 6 conducts the desired safety analysis. It verifies the design model to ensure the nominal behavior will not lead to the functional hazard, and analyzes the safety model to  demonstrate that the safety risk associated with the functional hazard is acceptable. 

\subsection{The core activities of MBSA}\label{sec:essential}
We further narrow down the scope of MBSA by identifying the core activities that an MBSA approach must conduct.

First, the main goal of the functional hazard identification of Arrow 1 and the casual factors identification of Arrow 2 is \emph{to identify}, which is exploratory in nature, and highly relies on the subject matter experts' knowledge. The model-based approach can provide a structured way to represent the identified hazards and causal factors, and systematically analyze the traceability between levels of abstraction, but not \emph{identify} them in the first place.
In fact, many MBSA approaches start with given hazards and/or causal factors. Therefore, we do not consider Arrow 1 and 2 as the core activities of MBSA.

Second, the characterization of human error of Arrow 3 has been traditionally addressed by the human factors community. Although system safety and human factors are closely related and there is a strong push to integrate human error modeling into safety assessment, at least at this point, the study of how humans make mistakes and how to model human error is still a separate topic from safety assessment. For example, in commercial aircraft certification, human factor (i.e., 25.1302) and system safety (i.e., 25.1309) still mostly work in parallel.
Therefore, we do not consider characterizing human error as a core activity of MBSA. 

Finally, the rest of the activities are the core activities of MBSA because no convincing case can be made for a safety analysis without explaining all the remaining activities no matter how narrow a view is taken for the safety analysis. In summary, the core activities of MBSA are the \textbf{component fault characterization} (Arrow 3), the \textbf{modeling language selection} (Arrow 4), the \textbf{safety model construction} (Arrow 5), and the \textbf{safety analysis} (Arrow 6).

Note that we are aware that some literature disagrees with this way of narrowing down the scope of MBSA. For example, \cite{woodham2018fueleap} puts the functional hazard identification under the umbrella of MBSA. However, 
one way to interpret our reasoning is to flip the logic. If one activity is an core MBSA activity, then any approach that does not conduct such activity is not MBSA. For example, many high-profile MBSA approaches do not specify how to model human error. If characterizing human error is a core activity, then all those well-known MBSA approaches will be disqualified as MBSA. For this reason, characterizing human error cannot be a core MBSA activity. The same reasoning process can be applied to other activities. As a caveat, although certain activities are not classified as the core activities of MBSA, that does not mean efforts should not be made to integrate them into MBSA. Excluding them from the core activities of MBSA is just to identify a \emph{minimal} scope of MBSA to reach a broad consensus across the community.

\section{The patterns of MBSA}\label{sec:feature}
In this section, we first classify the general patterns in which the core activities are conducted in the literature. After that, we summarize a minimal set of common characteristics of the general patterns (called the ``core patterns'') that an MBSA approach must exhibit.  

\subsection{Component fault characterization (Arrow 3 of Figure~\ref{fig:mbdsafety})}\label{sec:fault}
\subsubsection{Reasoning about component faults}\label{sec:fault-structure}
\begin{figure}[ht!]
    \centering
    \includegraphics[width=0.65\linewidth]{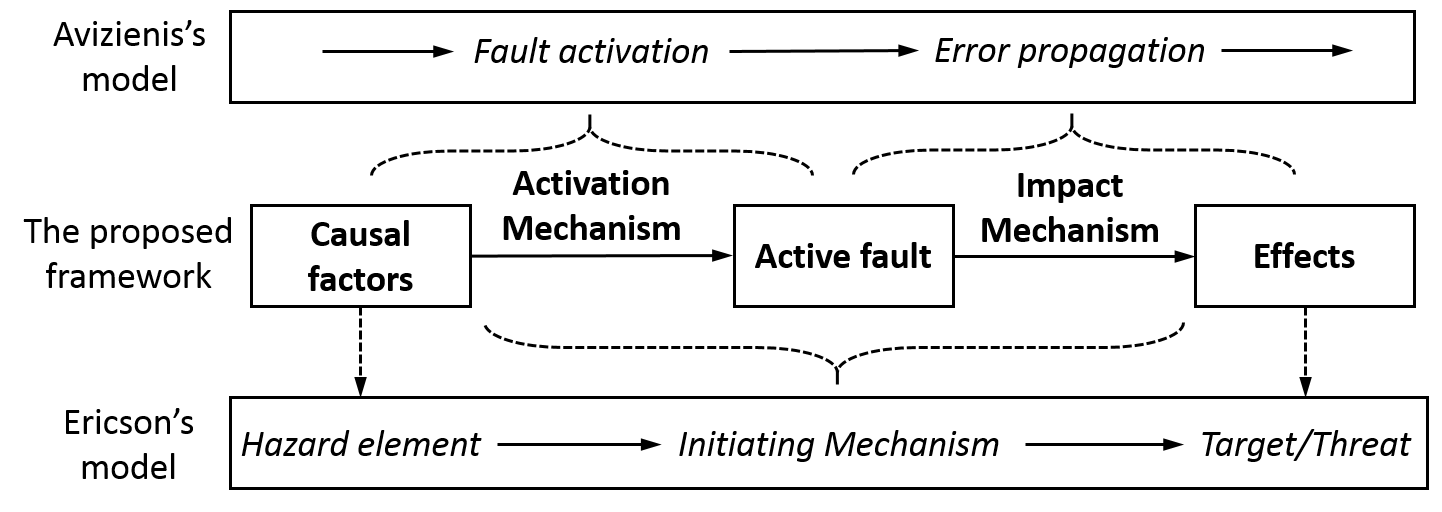}
    \caption{The proposed general framework of component faults, with Avizienis’s model \cite{avizienis2004basic} at the top and Ericson's model \cite{ericson2015hazard} at the bottom for comparison.}
    \label{fig:fault}
\end{figure}

Avizienis's model \cite{avizienis2004basic} on faults is a highly influential model in the dependable systems literature. It depicts how a fault activates and develops. According to Avizienis's model, fault activation is the application of the activation pattern to a component that causes a dormant fault to become active. Errors are produced by an active fault. Errors propagate to the system interface and cause the service delivered by the system to deviate from the correct service. There are two processes in Avizienis's model: an activation process that leads to an active fault, and an error propagation process that leads to the effects of the fault (i.e., the top of Fig. \ref{fig:fault}).

Another important work is Ericson's hazard model \cite{ericson2015hazard} (i.e., the bottom of Fig. \ref{fig:fault}). Ericson decomposes hazard into three elements: Hazardous Element, Initiating Mechanism and Target/Threat, where Hazardous Element is the basic hazardous resource creating the impetus for the hazard; Initiating Mechanism is the trigger or initiator event(s) causing the hazard to occur through the actualization or transformation of the hazard from a dormant state to an active mishap state; Target/Threat is the mishap outcome and the expected consequential damage and loss. 

Inspired by both Avizienis's model and Ericson's model, we propose a general framework to model component faults (i.e., the middle of Fig. \ref{fig:fault}). Specifically, each fault is correlated with a set of casual factors, all of which have to be present for a fault to happen or become active. Furthermore, a fault does not necessarily lead to a failure \cite{arp4754a2010guidelines} --- the notion of dormancy, which implies two mechanisms in play: one leading to the fault and the other leading to the effect of the fault. The two mechanisms are similar to the ``occurrence pattern'' and ``direct effect'' in \cite{ortmeier2007formal}. We denote them as the activation mechanism and the impact mechanism in this paper. Finally, effects are the outcomes of the fault, and there can be multiple ways to represent the effects of a fault. The relationship between the proposed framework and Avizienis's model and Ericson's model is depicted in Fig. \ref{fig:fault}.

\subsubsection{The general patterns of component fault characterization}
We examine each element in the framework proposed in Fig.~\ref{fig:fault} and classify the general patterns of component fault characterization in the literature. According to \cite{avizienis2004basic}, faults can be originated from the operational phase and developmental phase. However, faults modeled in MBSA literature are mostly faults at the operational phase.

\paragraph{Causal factors.}
While the specific context of causal factors vary from application to application, two perspectives are often presented in the literature about the causal factors: (1) the location of the causes with respect to the component boundary, and (2) the likelihood of the causes (i.e. the occurrence).

With respect to the \textbf{location}, there are internal causes and the external causes \cite{stewart2017architectural,bozzano2011safety,delange2014aadl,papadopoulos1999hierarchically}. The difference between the two is self-explanatory.

Moreover, the external causes can be further classified into intended and unintended. The intended external causes refer to the scenarios where the intended input becomes erroneous and leads to the fault. For example, a voltage (i.e. the input) too high can break a capacitor in the circuit. Examples abound in the literature \cite{boudali2008architectural,walter2008opensesame}.
The unintended external causes are relatively less studied by the model-based literature. They are
caused by the effect of a fault stemming from an external component that is not supposed to have interaction with the current component in nominal conditions. They are called ``unconnected'' in \cite{joshi2007behavioral}, and  ``unintended interaction faults'' in \cite{avizienis2004basic} . In fact, many causal factors addressed in Zonal Safety Analysis and Particular Risk Analysis \cite{arp47611996guidelines} belong to this type, such as engine burst and electromagnetic interference.

The \textbf{occurrence} of component fault can either be probabilistic or Boolean. For the internal causal factors, the occurrence usually follows a probability distribution. Industry has reliability dataset that records the probability distributions of the failures of commonly used components. For the external causal factors, probability can also be applied to describe the uncertainty that some external causes may or may not lead the component fault. When the causal factors are characterized with probability, both quantitative and qualitative safety analysis can be conducted \cite{gudemann2010framework}.

In addition, the occurrence of the component faults can also be depicted with Boolean variables. Safety analysis based on Boolean causal factors usually assumes the worst-case scenario and reaches conservative conclusions. Boolean causal factors are mainly used for qualitative safety analysis.

\paragraph{Activation mechanism.}
The causal factors being present is a necessary but not sufficient condition for the fault to be present. The activation mechanism is the additional condition  (also known as ``fault activation conditions'' in \cite{arlat1991fault}) that the causal factors must satisfy to activate the fault. We examined the literature and identified the following list of patterns for the activation mechanism.
\begin{itemize}
    \item Guard: The component may be vulnerable to the causal factor(s) only when it is in certain states. For example, overcurrent can cause an electric device to fail only when the device is on. 
    \item Sequence: The causal factors may have to happen in a certain sequence for the fault to be activated. This is one of the main advancements of Dynamic Fault Tree \cite{backstrom2016effective,junges2016uncovering} over the traditional FTA. In \cite{jayakumar2020property}, the fault activation conditions specify the input/state sequences that cause the activation of the fault and the propagation within the system.
    \item Delay: It may take time for a fault to happen even after the causal factors are all present. For example, a pump overheats after no water flows in for a certain period of time. This time period can be a deterministic one or a probabilistic one \cite{prosvirnova2013altarica}. If the fault is activated immediately after the causal factor(s) is present,  such a causal factor can usually be modelled as a trigger event of the fault \cite{kaiser2007state/event}.
    \item Duration: A duration may be defined to model the phenomenon that a fault can be deactivated. A fault can disappear a certain period of time after the activation because of its transient nature \cite{ortmeier2005deductive} or after being repaired \cite{chaux2011qualitative}. A infinite long duration means the fault is permanent. The characterization of the time can be deterministic, probabilistic \cite{ortmeier2005formal} or non-deterministic as suggested by \cite{gudemann2010probabilistic}. 
\end{itemize}

\paragraph{Impact mechanism.}
The fault being activated is a necessary but not sufficient condition for the fault to show effects on the component. The impact mechanism is the additional condition that needs to be satisfied for a given fault to cause the defined effects. We examined the literature
and identified the following list of patterns for the impact
mechanism.
\begin{itemize}
    \item Guard: The fault may lead to the defined effects only when the system is in certain state. For example, loss of hydraulic supply will only affect the ground deceleration function when the aircraft is landing.  Guard condition is widely modeled in the literature such as in state/event fault tree \cite{kaiser2007state/event} and AltaRica Data-flow \cite{seguin2004formal}. 
    \item Delay: As pointed of by \cite{lisagor2010failure}, ``the effect of a failure may not immediately cause an output failure mode and may remain dormant''. The time between the fault being present and the appearance of the fault effect is defined as ``Fault Tolerant Time Interval'' by \cite{gonschorek2019integrating}. This delay is considered as ``safety-relevant properties'' in SafeDeML \cite{gonschorek2019safedeml}.
    \item Indeterminism: The exact effects of a fault may be uncertain due to either epistemic uncertainty or aleatoric uncertainty. \cite{braman2009probabilistic} coins it as ``probabilistic transition conditions''. A feature called ``branching transition'' is provided by \cite{feiler2016architecture} where multiple target states can be transitioned to from one source state following a probability distribution. 
\end{itemize}

Note that guard and delay are defined in both the activation mechanism and the impact mechanism. Sometimes, they can even be modeled with the same modeling construct. However, they are inherently different in terms of their roles in a component fault. It is crucial to identify them correctly when characterizing the component faults and then model them accordingly with whatever modeling language selected. 

\paragraph{Effects on the component.}
Fault effects are characterized differently in the literature. Two dimensions are identified from the literature: the abstraction and the semantics (Table \ref{tab:effect}).

The abstraction dimension determines how many details are included in the effect model. At the \emph{component} level, how the fault affects the internal behaviors of a component is presented. At the \emph{architecture} level, the fault effects are only captured at the input/output ports of a component at the interface. In \cite{cuenot2014applying,fenelon1992new}, the former is called white box error model and the latter is called black box error model.
For the \emph{function} level, the component is abstracted as a Boolean logic node; the fault sets the node to be false, affecting all the functional flows that pass through the node. In fact, AADL-EMV2 supports modeling at exactly all the three levels of abstraction with a slightly different naming system \cite{kushal2017architecture,delange2014architecture}. Modeling at different levels of abstraction follows the safety analysis carried out (by techniques such as FMEA and HAZOP) at various stages of the system development life cycle. At the concept phase, the safety analysis is congruent to the level of abstraction of the functional model. At the system level and detailed design stage, safety analysis is congruent to the abstraction at the architecture and component level respectively.

The semantics dimension classifies different ways to interpret the effect of a fault. First, local effect of the fault can be interpreted as a deviation to the intended performance of the nominal behavior. In this way, the semantics of the \emph{nominal} behavior can be reused, and the global effects of the fault can be evaluated using the same interaction paths of the nominal behavior.
Second, the local effect of a fault can also be interpreted as new off-nominal behaviors (such as omission, commission, early and late)
beyond the semantics of the nominal behavior. Depending on how the off-nominal behaviors of components downstream are triggered, there are two different ways to evaluate the global effects. On one hand, the downstream off-nominal behavior can be triggered directly by the upstream off-nominal behavior. In this case, the safety model only includes the \emph{off-nominal} behavior. 
On the other hand, the downstream off-nominal behavior can be modeled as a result of the interactions between the upstream off-nominal behavior and the downstream nominal behavior. In this case, both nominal behaviors and off-nominal behaviors are present in the safety model (thus \emph{integrated}). 

In fact, the semantics dimension is consistent with the classification of FEM and FLM in \cite{lisagor2011model-based}. We classify the effect patterns of a sample pool of literature in Table~\ref{tab:effect}. 
\begin{table}
\caption{Patterns that a sample pool of literature model the fault effects on a component. Note that some papers appear in more than one cell because the fault effects are defined at multiple levels of abstraction.}
\label{tab:effect}
\begin{tabularx}{\textwidth}{|X|X|X|X|}
\hline
 \backslashbox{\textbf{Abstraction}}{\textbf{Semantics}}& \textbf{Nominal} &  \textbf{Off-nominal} & \textbf{Integrated} \\ \hline
\textbf{Function}     & NA               & \cite{kaiser2003new,helle2012automatic}       & \cite{delange2014aadl,kaiser2007state/event,boudali2008architectural} \\ \hline
\textbf{Architecture} & \cite{bonfiglio2015software, dong2019model-based,joshi2005model-based}&   \cite{chaari2016transformation,kaiser2003new,fenelon1992new,gonschorek2019integrating,papadopoulos2013hip,wallace2005modular,yang2019fds-ml,gomes2012constructive,bozzano2014formal}  & \cite{prosvirnova2014altarica,delange2014aadl,kaiser2007state/event,boudali2008architectural,bozzano2011safety,seguin2004formal,leitnerfischer2011quantitative,zhao2016failure,liu2012functional} \\ \hline
\textbf{Component}    & \cite{bozzano2003improving,joshi2007behavioral,stewart2017architectural,bretschneider2004model,peikenkamp2006towards} & \cite{gonschorek2019integrating} & \cite{prosvirnova2014altarica,delange2014aadl,kaiser2007state/event,bozzano2011safety,seguin2004formal,leitnerfischer2011quantitative,zhao2016failure}\\ \hline
\end{tabularx}
\end{table}

\begin{figure}[ht!]
    \centering
    \includegraphics[width=0.99\linewidth]{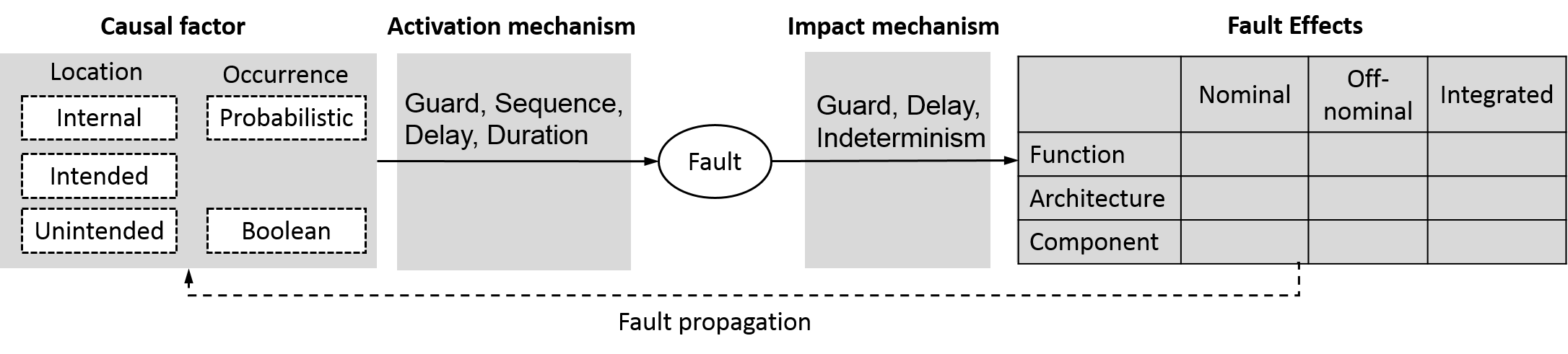}
    \caption{The general patterns of component fault characterization. }
    \label{fig:unifying}
\end{figure}

\paragraph{Summary.} As a result, the general patterns of the component fault characterization can be summarized in Fig.\ref{fig:unifying}. Note that when one identifies the fault of a component as the external casual factor of another component, the fault propagation between the two components will be automatically defined after the fault effect of the upstream component and the activation mechanism of the downstream component are defined. For this reason, we do not define a dedicated fault propagation mechanism to avoid repetition. 

\subsection{Modeling languages selection (Arrow 4 of Figure~\ref{fig:mbdsafety})}\label{sec:modeling}
\subsubsection{Reasoning about how the modeling language is selected}\label{sec:select}
\begin{figure}[ht!]
    \centering
    \includegraphics[width=0.3\linewidth]{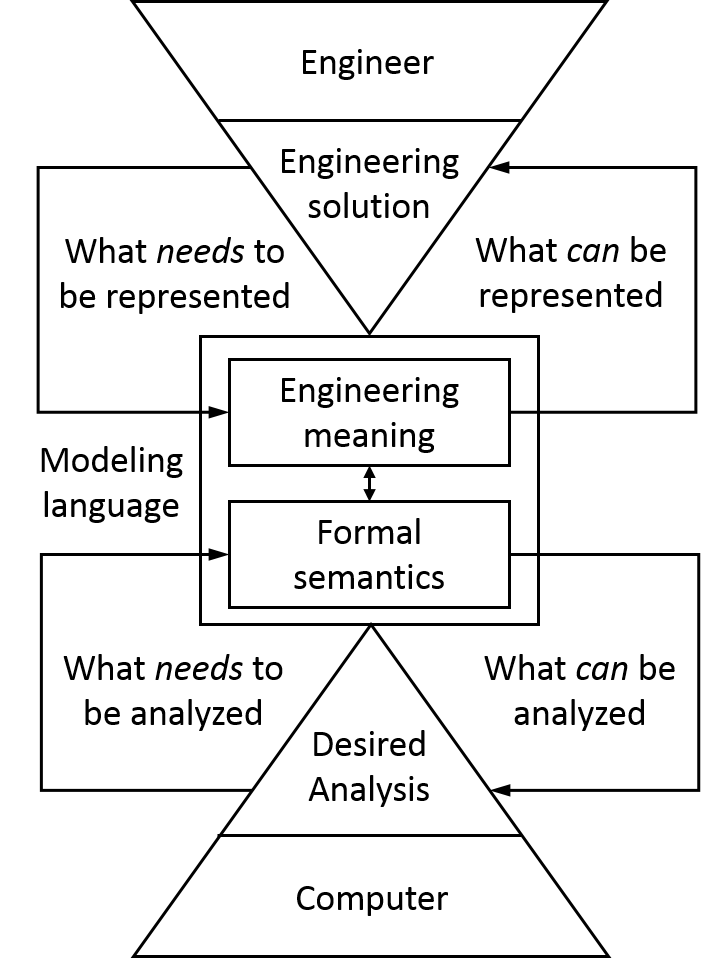}
    \caption{The modeling language must be able to fully represent the engineering solution, and must be analyzable by a computer program for the desired analysis.}
    \label{fig:language}
\end{figure}

A model in the context of MBD (Fig.\ref{fig:mbd_process}) is intended to (1) represent the engineering solution, and (2) facilitate the desired analysis by a computer program. As shown in Fig.\ref{fig:language}, the engineering solution specifies what \emph{needs} to be represented, and the desired analysis specifies what analysis \emph{needs} to be conducted. However, a specific modeling language has a limited capability of what \emph{can} be represented, and what analysis \emph{can} be conducted. Therefore, the modeling language must support both the modeling efforts (i.e., be fully equipped to represent the engineering solution) and the computing efforts (i.e., be computable by a computer program for the desired analysis), which in fact are the two challenges  observed in \cite{france2007model-driven}: ``the abstraction challenge'' and ``the formality challenge''.

Furthermore, the modeling efforts assign engineering meanings to the syntax of the language, and the computing efforts map the engineering meanings (through the syntax) into the formal semantics for compatible programs to compute. Therefore, \emph{engineering meaning} and \emph{formal semantics} (Fig.\ref{fig:language}) are the two inherent aspects of modeling languages in MBD. 
For example, a Simulink block has a pre-defined engineering interpretation at the front end to represent engineering components (e.g., a high-pass filter) and a formal semantics at the back end to support the mathematical analysis.
In addition, ``$\int$'' in an engineering solution can be the relationship between the acceleration and the velocity; it can also be the mathematical operation of ``integral'' in a computer program. Therefore, modeling languages in MBD always have the two aspects.
Similar concepts can be found as in ``pragmatic vs. formal'' of \cite{batteux2019model}, and ``interpretation vs. theory'' of \cite{seidewitz2003models}.

\subsubsection{The general patterns of the modeling languages}\label{sec:language}
We have identified the two aspects of a modeling language: engineering meaning and formal semantics. 
Different modeling languages have different emphases. Some focus on supporting the modeling efforts, some are designed mainly to compute properties, and some address both. Therefore, based on how the engineering meaning and the formal semantics are specified in a modeling language, we classify the safety modeling languages from the literature into the following three classes. 
\begin{itemize}
    \item Modeling-oriented: Modeling languages in this class are developed mainly to represent the system (off-nominal) behavior. Usually, no dedicated formal semantics is specified for the language. To use these languages for safety assessment, safety engineers have to make the computing efforts either by designing tools from scratch, or by transforming the language to existing and well-supported formalism (or languages), in order to conduct the desired safety analysis, such as UML in \cite{leitnerfischer2011quantitative}, Sysml in \cite{yakymets2015model-driven,david2010reliability}, HipHops in \cite{feiler2017automated} and AADL in \cite{rugina2007system, mokos2010ontology}.
    \item Computing-oriented: Modeling languages in this class are very close to mathematical formalisms. The engineering concepts under study are usually represented ad-hoc, as the mathematical objects can always be assigned with different engineering meanings \cite{rauzy2019foundations}. Safety engineers define the off-nominal behavior directly using the mathematical construct such as the Interface Automaton in \cite{zhao2016failure} and Hybrid Automaton \cite{liu2012functional}. Usually, this type of modeling languages is well-supported by existing tools for the desired safety analysis. 
    \item Comprehensive: Modeling languages in this class usually have syntax that is associated with certain engineering meanings (e.g., error propagation), and a formal semantics, for example, SMV \cite{bozzano2007fsap} and Statemate \cite{bretschneider2004model,peikenkamp2006towards} with finite state machine, SLIM with Event Data Automation \cite{bozzano2011safety}, AltaRica with General Transition System \cite{batteux2019modeling} and Arcade with Input/output interactive Markov chains \cite{boudali2008architectural}. 
\end{itemize}

Five criteria are developed in \cite{boudali2008arcade-a} to qualitatively compare the effectiveness of a modeling language in terms of the modeling efforts and the computing efforts. In our case, the modeling-oriented languages have the best support for the modelling efforts, but require considerable efforts to accomplish an automatic computerized safety analysis. The computing-oriented languages have the best support in conducting automatic safety analysis, but can be challenging when modeling a complex system. The comprehensive languages provide support in both modeling and computing. 

\subsection{Safety model construction (Arrow 5 of Figure~\ref{fig:mbdsafety})}
\subsubsection{Reasoning about how safety model is constructed}
This activity constructs the safety model from the design model. Making sure the safety model is consistent with the design model has always been a challenge in the system safety community. 

The design model represents the nominal behavior of a system. While how a fault develops within a component has no logical consistency with the nominal behavior of the component, how the fault may propagate across the system does bear (at least partially) resemblance with the interaction path of the nominal behavior. Such resemblance is indeed the basis for the consistency between the design model and the safety model. 
 Because the fault propagation path and the nominal interaction path are usually considered as the architectures of the safety model and the design model respectively, the consistency between the two models can be represented as the consistency between the architectures, which we call \textbf{architecture consistency} in this paper. 
 
 In other words, regardless of how the models are constructed, the architecture consistency between the design model and the safety model must be maintained, which is, in fact, one of the most recognized features of MBSA \cite{joshi2005proposal,piriou2014control,frazza2022mbsa}. Compared to the traditional safety approaches, such as FTA, that are not formally related to the technical architecture of the system \cite{kaiser2018advances}, the MBSA models represent systems with a point of view closer to their architecture \cite{vidalie2022category}. 
 Some literature even addresses the MBSA approaches as ``architecture-based safety evaluation methodologies'' \cite{grunske2008comparative}.

\subsubsection{The general patterns of architecture consistency}
\begin{figure}[htb]
    \centering
    \includegraphics[width=0.59\linewidth]{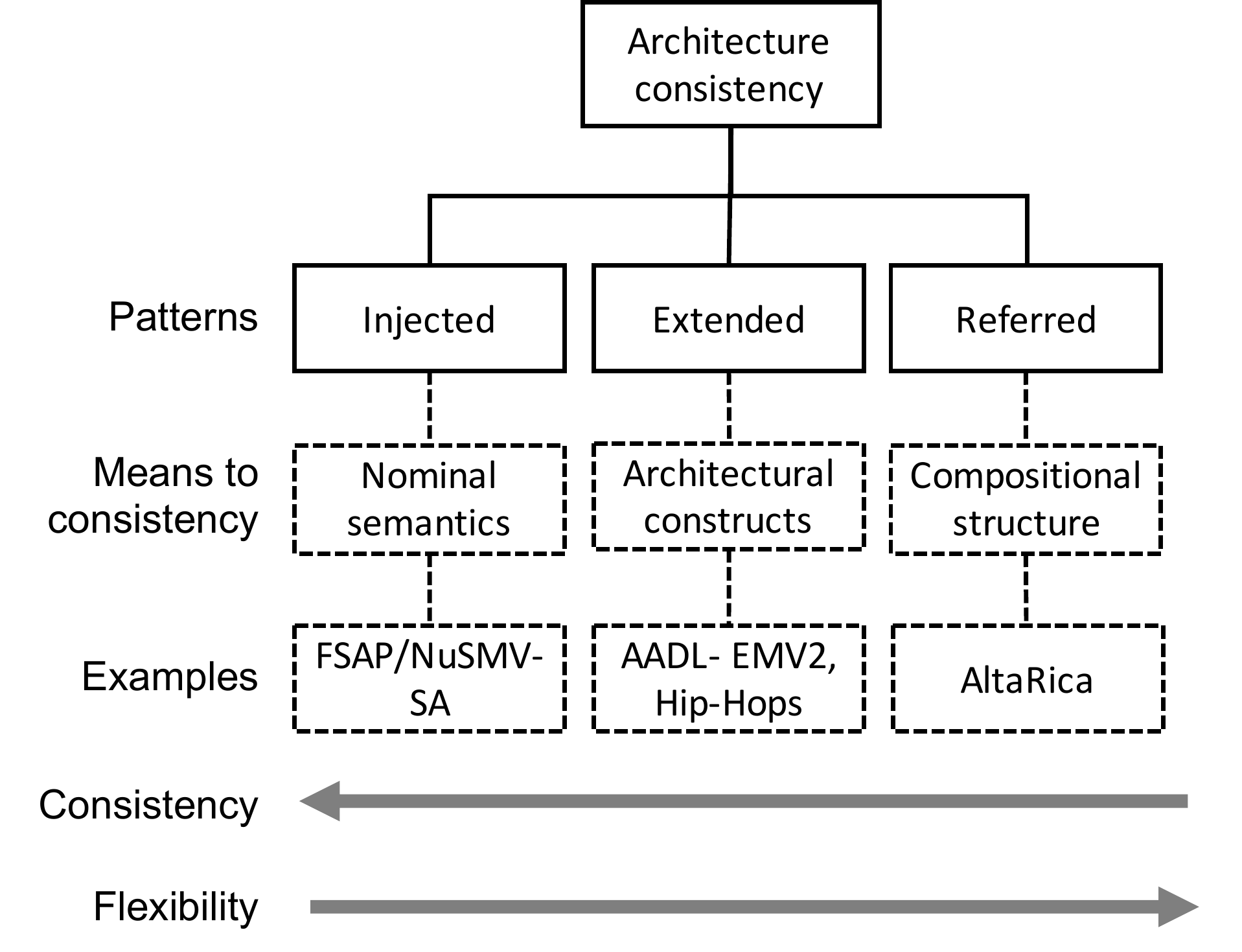}
    \caption{Different patterns that the architecture consistency is achieved in the literature.}
    \label{fig:ac}
\end{figure}

Three different ways of achieving Architecture Consistency is found in the literature (Fig.\ref{fig:ac}). This classification is an extension of the model provenance in \cite{lisagor2011model-based} and the ESACS project in \cite{bozzano2003esacs,bozzano2003improving1}. 

The first class injects faults into the design model. The off-nominal behavior of the individual component is described by reusing the semantics of the nominal behavior, and then is ``injected'' into the design model. No information of the fault propagation paths is defined. Instead, the fault propagation paths are generated automatically by reusing the interaction paths of the nominal behavior, which leads to the architecture consistency of the design model and the safety model by construction. 
Therefore, the architecture consistency in this class is achieved by reusing the \emph{nominal semantics}.
A typical example is the FSAP/NuSMV-SA language \cite{bozzano2007fsap}.

The second class extends the design model into to the safety model by reusing the \emph{architectural constructs} of the design model when defining the off-nominal behaviors of each component. Tools are usually available to ensure the architecture consistency between the design model and the safety model through the shared architectural constructs. Typical example is AADL- EMV2 \cite{delange2014aadl} and HiP-HOPS for EAST-ADL \cite{chen2013systems}.

For the third class, the safety engineers refer the \emph{compositional structure} of the design model to manually create the safety model. The safety model is organized and constructed in a compositional way \cite{parker2013model} so that the architecture consistency can be examined by comparing the compositional structure of the design model and the safety model. 
Typical examples include AltaRica \cite{prosvirnova2014altarica}, Component Fault Tree \cite{kaiser2003new}, Failure Propagation and Transformation Notation \cite{mcdermid1995experience}, Fault Propagation and Transformation Calculus \cite{fenelon1992new} and State Event Fault Trees \cite{kaiser2007state/event}.

\paragraph{Observation.} 
From left to right in Fig.\ref{fig:ac}, there is less and less support in maintaining architecture consistency between the design model and the safety model. This is the same argument as ``model provenance'' in \cite{lisagor2011model-based} and we will not provide further explanation here. 

However, the flexibility to model complex off-nominal behaviors goes in the opposite direction. For the ``injected'' class, faults are depicted as the deviation from the nominal behavior. But it is possible that certain off-nominal behaviors caused by the faults cannot be expressed as deviation of the original design model. Therefore, the ``injected'' way to achieve architecture consistency has limited ability to express complex off-nominal behaviors.  
By contrast, both the ``extended'' class and the ``referred'' class have dedicated semantics to represent the off-nominal behaviors, and that provides more flexibility in describing the complex off-nominal behaviors because the safety engineers do not have to be restricted by the nominal semantics of the design model. 

Furthermore, the off-nominal behaviors of a component usually evolve from the nominal behaviors of the same component. The ability to depict this evolution process will make the safety model more accurate in representing the complex off-nominal behaviors. 
In this regard, the ``referred'' class works on a ``clean slate'', and hence has the freedom to make abstractions of the nominal behavior when describing the evolution process aforementioned, while the ``extended'' class is (at least partially) constrained by the existing design model for the same endeavor. For this reason, the ``referred'' class has more flexibility than the ``extended'' class in describing the complex off-nominal behaviors.
Therefore, as shown in Fig.\ref{fig:ac}, from left to right, it becomes more and more flexible to describe complex off-nominal behaviors. 

In fact, architecture consistency is a type of architectural constraint on the safety model. The weaker constraints on the architecture of the safety model, the more flexible the safety model is able to describe complex off-nominal behaviors, which explains the opposite directions of the ``consistency'' and the ``flexibility'' in Fig.\ref{fig:ac}.

\subsection{Safety analysis (Arrow 6 of Figure~\ref{fig:mbdsafety})}
\subsubsection{Reasoning about how safety is analyzed}
\begin{figure}[htb]
    \centering
    \includegraphics[width=0.27\linewidth]{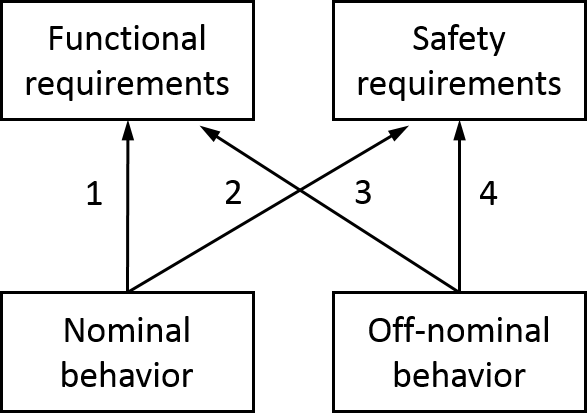}
    \caption{Four types of safety analysis.}
    \label{fig:safety_analysis}
\end{figure}

In general, safety analysis demonstrates that system behaviors satisfy requirements. The system behaviors can be nominal or off-nominal, and the requirements can be the functional requirements or the safety requirements. In this way, we can derive four general types of safety analysis in Fig. \ref{fig:safety_analysis}.

\begin{itemize}
    \item Type 1 analysis demonstrates the ``goodness'' of the design. The nominal behaviors have to satisfy the functional requirements, which is the foundation of all safety analysis. 
    \item Type 2 analysis demonstrates that the nominal behavior will satisfy the safety requirements. For example, a series of safety assessment is conducted in \cite{althoff2010reachability} to prove that all the possible trajectories of the autonomous cars are not in conflicts by using reachability analysis. 
    \item Type 3 analysis demonstrates the ``fail-operational'' \cite{sinha2011architectural} property, i.e. the functional requirements can still be satisfied even in the presence of component failure. This is a subject of robustness analysis \cite{krach2017model-based}, a dependability property that is closely related to safety. 
    \item Type 4 analysis demonstrates the fail-safe property, i.e., the system is still safe even in the presence of component failure.
\end{itemize}

\subsubsection{The general patterns of safety analysis}
In this section, we provide a list of the different types of safety analysis found in the literature. Note that only the commonly conducted safety analysis is listed here. The less common analysis, such as the safety optimization in \cite{papadopoulos2013hip}, is not included here.  
\begin{itemize}
    \item [(1)] Fault tree analysis: This class of analysis includes the automatic generation of fault tree, the derivation of the minimal cut sets and the calculation of the failure rate \cite{gradel2022model}. Note that not all approaches conducts all the three tasks. For example, \cite{kaiser2003new} only mentions the generation of the fault tree; \cite{yang2019fds-ml} only derives the minimal cut set; \cite{gomes2012constructive} only addresses the failure rate calculation for the hazard. However, because all the three activities are part of a conventional FTA, they are all classified in the class of fault tree analysis. 
    \item [(2)] Failure modes and effects analysis (FMEA): This class of analysis is the automatic generation of FMEA \cite{adedjouma2019framework,bonfiglio2015software}.
    \item [(3)]  Reliability Block Diagram (RBD): This class of analysis is similar with FMEA and can be automatically accomplished, for example in \cite{delange2014aadl} and \cite{helle2012automatic}. 
    \item [(4)] Probabilistic indicators: This class of analysis is a broad class of analysis for dependability, such as Reliability and Availability. We direct readers to \cite{avizienis2004basic} for a detailed explanation. Example approaches that cover analysis of probabilistic indicators include \cite{bozzano2011safety} and \cite{dong2019model-based}.
    \item [(5)] Property verification for nominal behavior: This class of analysis is the formal verification of nominal behavior (e.g. \cite{stewart2017architectural,joshi2005model-based}) against the function and/or the safety properties. It checks the ``goodness'' of a design, which is the precondition for any safety analysis.
    \item [(6)] Property verification for off-nominal behavior: This class of analysis is the formal verification of off-nominal behavior (e.g. \cite{bozzano2015formal} and \cite{liu2012functional}) against safety properties. It rigorously checks whether certain safety constraints can be broken under certain off-nominal conditions.
    \item [(7)] Critical sequence: A critical sequence is a sequence of events leading from the initial state to a critical state. In the case of dynamic models, the order of occurrences of events is important and thus a simple minimal cut set is insufficient: minimal or most probable sequences or sequences of a given length (also called order) can be extracted by simulation of the model. Example works include \cite{prosvirnova2014altarica,leitnerfischer2011quantitative,leitner-fischer2011quantum:,beer2012analysis}.
    \item [(8)] Trace simulation: This class of analysis is to display the traces of the individual failure scenarios for the safety engineering to debug the model and understand the propagation of the fault effect. It can be a step-wise interactive simulation \cite{seguin2004formal} or the trace of a given number of steps is output at the end of the simulation \cite{bozzano2007fsap}.
    \item [(9)] Common cause analysis: This class of analysis is required in \cite{arp47611996guidelines} and aims at investigating possible dependencies between the faults and evaluates the consequences in terms of system safety/reliability \cite{kessler2019}. Example works include \cite{bittner2016xsap,peikenkamp2006towards}.
\end{itemize}

\subsection{The core patterns of MBSA}\label{sec:mbsa}
We summarize the general patterns in which the core activities are conducted in Table. \ref{tab:patterns}. 
Among the general patterns, we identify in this section a minimal set of common characteristics as the core patterns that an MBSA approach must exhibit.
\begin{table}[]
\caption{The general patterns of MBSA.}
\label{tab:patterns}
\begin{tabular}{|l|l|lll|}
\hline
   \textbf{Arrow NO.}               &     \textbf{Activity}              & \multicolumn{3}{l|}{\textbf{The general patterns}}                                             \\ \hline
\multirow{6}{*}{3} & \multirow{6}{*}{Component fault characterization} & \multicolumn{1}{l|}{\multirow{2}{*}{Causal factor}} & \multicolumn{1}{l|}{Location} & Internal, Intended, Unintended  \\ \cline{4-5} 
                  &                   & \multicolumn{1}{l|}{}                  & \multicolumn{1}{l|}{Occurrence} & Probabilistic, Boolean \\ \cline{3-5} 
                  &                   & \multicolumn{2}{l|}{Activation Mechanism}                                          & Guard, Sequence, Delay, Duration \\ \cline{3-5} 
                  &                   & \multicolumn{2}{l|}{Impact Mechanism}                                          & Guard, Delay, Indeterminism \\ \cline{3-5} 
                  &                   & \multicolumn{1}{l|}{\multirow{2}{*}{Effect}} & \multicolumn{1}{l|}{Abstraction} & Function, Architecture, Component \\ \cline{4-5} 
                 &                 & \multicolumn{1}{l|}{}                  & \multicolumn{1}{l|}{Semantics} & Nominal, Off-nominal, Integrated \\ \hline
                  4 &      Modeling languages selection             & \multicolumn{3}{l|}{Modeling-oriented, Computing-oriented, Comprehensive}                 \\ \hline
                 5 &      Safety model construction             & \multicolumn{2}{l|}{Architecture Consistency} &Injected, Extended,Referred                    \\ \hline
                 6 &      Safety analysis             & \multicolumn{3}{l|}{The nine types of safety analysis}                \\ \hline
\end{tabular}
\end{table}

\paragraph{Component fault characterization (Arrow 3).} Almost all the literature characterizes component faults and the fault propagation between components. Compared to the traditional approaches like FTA and FMEA where faults are only abstracted as Boolean variables, the model-based approaches are able to capture more sophisticated dependencies between the components and obtain more precise analysis results \cite{gudemann2010quantitative}.

\paragraph{Modeling languages selection (Arrow 4).} The commonalities between the modeling languages used in different literature are that they all can (1) represent the component faults and (2) support automatic safety analysis.

\paragraph{Safety model construction (Arrow 5).} Not all literature emphasizes architecture consistency between the safety model and the design model. However, almost all the prominent approaches implement architecture consistency as one of the key missions of MBSA. Architecture consistency is indeed an important constraint for any safety analysis. Therefore, we consider architecture consistency as a common practice among the literature.

\paragraph{Safety analysis (Arrow 6)} Among the four types analysis in Fig.\ref{fig:safety_analysis}, almost all the literature conduct at least Type 4 analysis. The focus of classical safety analysis techniques lies on supporting the reasoning of possible failures and the causal relationships in failure events \cite{cuenot2007towards}. Therefore, demonstrating the system is safe in the presence of component failure is a minimal commonality of the literature. 

\vspace{2mm}
\noindent Therefore, the \textbf{core patterns} of MBSA are,
\begin{itemize}
    \item It must model component fault and fault propagation.
    \item It must support automatic safety analysis. 
    \item It must maintain architecture consistency between the design model and the safety model.
    \item It must evaluate whether the system is safe in presence of component failure. 
\end{itemize}

\section{A case study}
In this section, we review a recent MBSA paper \cite{stewart2021aadl} (addressed as the \aadl) to test the effectiveness of the characterization of MBSA in this paper.
Note that we intentionally excluded the \aadl when we characterized the MBSA activities and patterns so that the results would not be biased toward the specific practices of the \aadl. 

\paragraph{Component fault characterization (Arrow 3).}
The \aadl explains in its Section 3.6 how to model the individual faults and the fault propagation between the components, which apparently satisfies the core pattern for this activity. Specifically, 
\begin{itemize}
    \item Causal factor: The \aadl addresses two types of causal factors. First, they described an internal causal factor leading to ``inverted\_fail'' with probability $5.0\times 10^{-6}$. Second, they mentioned ``a failure in one hardware component may trigger failure in other hardware components located nearby'', which is in fact an ``unintended'' external causal factor. Of note, they also mentioned the faulty behavior of one component may lead to a violation of the contracts of other components in the downstream components. But it is not clear whether such a violation will cause a fault of the downstream component. If yes, that would be an `intended'' external causal factor.  
\item Activation mechanism: Not much was defined in the \aadl  for the activation mechanism except the ``duration'' of the fault.
\item Impact mechanism: The \aadl discussed different destination components may observe different effects of the same component fault (i.e., the asymmetric fault). This phenomenon can be easily modeled by defining respective impact mechanisms for different destination components. Furthermore, non-deterministic communication nodes are used to model this phenomenon, which is also aligned with the ``indeterminism'' of our impact mechanism. 
\item Effect model: First, once a fault is activated, it ``modifies the output of the component''. Hence, the \aadl models the fault effects at the output port of a component, i.e., the ``architecture'' at the abstraction dimension. Second, because ``the faulty behavior propagates through the nominal behavior contracts in the system model just as in the real system'', it belongs to the ``nominal'' class at the semantics dimension. 
\end{itemize}
\paragraph{Modeling language selection (Arrow 4).}
The \aadl explains in its Section 3.7 how the safety model is analyzed formally by the model checker JKind.  
Apparently, the automatic safety analysis is supported, which satisfies the core pattern of this activity. 

Specifically, the AADL/AGREE language is modeling-oriented, because the compositional nature of the modeling language makes it straightforward to model a complex system hierarchically. In addition, to facilitate the automatic safety analysis, the AGREE tool translates the language into Lustre, and then to the model checker JKind. Such a series of translation steps is an important feature of the modeling-oriented language, as explained in Section \ref{sec:language}. 
\paragraph{Safety model construction (Arrow 5).} One of the key objectives of the \aadl is to ``support a shared model'' between the design side and the safety side. 
An considerable amount of efforts were made to emphasize that faults in the \aadl were injected (or the ``implicit error propagation'') into the the design model. Clearly, the \aadl satisfies the core pattern for this activity and achieves architecture consistency in the ``injected'' way.   
\paragraph{Safety analysis (Arrow 6).} 
It was explicitly mentioned in the \aadl that the safety analysis ``verify safety properties in the presence of faults'', which satisfies the core pattern for this activity. Furthermore, the \aadl checked the design model against the safety properties, generated the minimal cut set and calculated the probability of combinations of faults, which corresponds to (1) and (5) of the safety analysis. 

\paragraph{Summary.} The \aadl conducts all the core activities of MBSA and exhibits all the core patterns of MBSA. None of the non-core activities (i.e., functional hazard identification, causal factor identification and human error characterization) are conducted in the \aadl. The specific patterns in which the \aadl conducts the core activities are also captured in the general patterns characterized by our paper. The results indicate that our characterization of MBSA is effective to reason about MBSA.

\section{Conclusion}\label{sec:conclusion}
\subsection{Summary}
There is no consensus on the identity of MBSA, which impedes the advancement of MBSA. We investigated the identity of MBSA in this paper. 
First, we reasoned about the core activities that an MBSA approach must conduct. Second, we characterized a set of general patterns in which the core activities are conducted in the literature. After that, we summarized a minimal set of common characteristics of the general patterns (i.e., the core patterns) that an approach must exhibit to be considered MBSA. Finally, we successfully tested the effectiveness of the proposed characterization of MBSA on a recent MBSA paper. 

We welcome other and potentially different characterizations of MBSA, as MBSA is just a way to conduct safety analysis, and there should be no standard approach to it.
We hope this analysis of MBSA will eventually lead to a general consensus on the identity of MBSA, promoting the advancement of MBSA as a leading research topic in the system safety engineering community.

\subsection{Suggestions}
Over a decade ago, Lisagor observed that most of the MBSA innovation focuses on model specification notations and/or algorithms for possible manipulations of the models \cite{lisagor2010illusion}. Braun \cite{braun2009model} pointed out that the major open issue is how to reason about the choice of models. Most recently, Sadeghi \cite{sadeghi2021state} made a similar observation that ``establishing the validity of the models (of MBSA) is still a major challenge''.  
Our investigation concurs with all these findings and identifies three reasons (and thus research opportunities) why there is a lack of attention to model validity in MBSA. 

\paragraph{Identifying faults.}
The safety model is invalid if some of the faults that may contribute to the hazards are missing. 

Finding a complete set of fault modes for a given component is not an easy task \cite{ortmeier2006failure-sensitive}. The combinatorial diversity of each plausible (fault) event interacting with other set of events within and without the system makes bottom-up analysis intractable, so experts' knowledge of the system behavior needs to be employed to narrow the search space of critical scenarios \cite{wille2019contributions}.
Traditional FTA relies on experts' knowledge to identify the causal factors of the failure event deductively.
However, the automatic FTA \cite{dickerson2018formal,clegg2019integrating,bozzano2007symbolic} in most MBSA approaches automates away the deductive process of the traditional FTA without proposing a replacement.
Although many MBSA approaches are supported by formal methods, which are known for the ability of exhaustive exploration, ``if a failure mode is not even part of the formal model, then it is impossible to reason about it'' \cite{ortmeier2006failure-sensitive}.

This general lack of deductive analysis in the MBSA literature is perhaps caused by the faster advancement of  formal methods than safety engineering (to which the deductive safety analysis is almost exclusively belong). Moving forward, it is crucial for the safety engineering community to develop new or modify existing deductive analysis techniques to integrate with the current MBSA approaches, so that the faults that may contribute to the hazards can be thoroughly identified in a more systematic, tractable and efficient way. 

\paragraph{Fault behavior modeling.}
The safety model is invalid if some of the faults are characterized incorrectly in the model. 

Unlike other scientific communities such as aerodynamics and heat transfer where the fundamental attributes of the subject phenomenon are well studied, the safety community only has limited studies on the general mechanisms of how a component fails. 
As a result, safety engineers decide the necessary information to characterize a component fault in an ad-hoc way, which makes it challenging to assure the adequacy of the fault models, especially for the newly designed components.  

Therefore, more efforts need to be made to study component faults as a scientific phenomenon, to help safety
engineers better understand the fault behaviors under study and therefore better characterize them in the safety model.
The fault patterns characterized in Fig.~\ref{fig:unifying} are our first attempt towards this goal. They constitute a generic model of
component faults, and should be continuously improved as more literature is reviewed.

\paragraph{Hazard without failure.} The safety model is invalid if the non-failure causal factors are not eliminated from the design model in the first place.

It has been well established that hazards can be caused by non-failure factors \cite{fleming2013safety}. However, the current MBSA
practices are overwhelmingly failure-oriented. Although EAST-ADL framework mentioned systematic faults such as
design errors \cite{chen2011integrated}, it is unclear from the literature how the design errors are actually identified in the first place.
Furthermore, it is also well established that (formal) verification alone is insufficient to validate safety \cite{butka2015advanced,leveson2020you}. Even
if the design model is formally verified, there is still no guarantee that the design is free from the non-failure causal
factors if certain safety-critical scenarios are missing from the design model.

Therefore, more research is needed on how to eliminate the non-failure causal factors from the design model. STAMP-STPA \cite{leveson2016engineering} is known for its ability to address the non-failure causal factors. Based on STPA, a new approach called STPA+ is proposed in \cite{sun2022new} to complement MBSA specifically for the non-failure causal factors. However, studies on this front are still rare in the safety communities. Without demonstrating that the non-failure causal factors are sufficiently addressed, the results of an MBSA approach still cannot be fully trusted for safety assurance.

\bibliographystyle{cas-model2-names}

\bibliography{mbsa}


\end{document}